\input{aipcheck}

\documentclass[
    ,sort
    ,final            % use final for the camera ready runs
  ]
  {aipproc}

\layoutstyle{8x11single}

\usepackage{amsmath}
\usepackage{wrapfig}

\newcommand{\gf}{\gamma_5}

\newcommand{\bra}{\langle}
\newcommand{\ket}{\rangle}
\newcommand{\braket}[1]{\bra #1 \ket}

\newcommand{\xx}{\braket{x^2}}

\makeatletter
\def\simleq{\mathrel{\mathpalette\gl@align<}}
\def\simgeq{\mathrel{\mathpalette\gl@align>}}
\def\gl@align#1#2{\lower.6ex\vbox{\baselineskip\z@skip\lineskip\z@
     \ialign{$\m@th#1\hfill##\hfil$\crcr#2\crcr\sim\crcr}}}

\makeatother

\begin{document}

\title{Nucleon strangeness form factors and moments of PDF}

%\classification{}

\author{Takumi~Doi}
{
  address={
Graduate School of Pure and Applied Sciences,
University of Tsukuba,
%Tennodai 1-1-1,
Tsukuba, Ibaraki 305-8571, Japan
email: doi@ribf.riken.jp
},
% ,email={doi@ribf.riken.jp}
}

\author{Mridupawan~Deka}{
  address={
Institute of Mathematical Sciences, 
Chennai- 6000113, India
}
}

\author{Shao-Jing~Dong}
{
  address={
Department of Physics and Astronomy,
University of Kentucky, Lexington KY 40506, USA}
}

\author{Terrence~Draper}
{
  address={
Department of Physics and Astronomy,
University of Kentucky, Lexington KY 40506, USA}
}

\author{Keh-Fei~Liu}
{
  address={
Department of Physics and Astronomy,
University of Kentucky, Lexington KY 40506, USA}
}

\author{Devdatta~Mankame}
{
  address={
Department of Physics and Astronomy,
University of Kentucky, Lexington KY 40506, USA}
}

\author{Nilmani~Mathur}
{
  address={
Department of Theoretical Physics,
Tata Institute of Fundamental Research,
Mumbai 40005, India}
}

\author{Thomas~Streuer}
{
  address={
Institute for Theoretical Physics,
University of Regensburg, 93040 Regensburg, Germany}
}

%\author{for $\chi$QCD Collaboration}{
%  address={ }
%}

%\author{<author3>}{
%  address={<common address for author2 and author3>}
%  ,altaddress={<author1 address>} % additional visiting address
%}

\begin{abstract}

The calculation of the nucleon strangeness form factors
from $N_f=2+1$ clover fermion lattice QCD
is presented.
Disconnected insertions are evaluated
using the Z(4) stochastic method,
along with unbiased subtractions from the hopping parameter expansion.
We find that increasing the number of
nucleon sources for each configuration improves the signal significantly.
We obtain $G_M^s(0) = -0.017(25)(07)$, 
%where the first error is statistical,
%and the second is the uncertainties in $Q^2$ and chiral extrapolations.
which is consistent with experimental values,
and has an order of magnitude smaller error.
Preliminary results for the 
strangeness contribution to the 
second moment of the parton distribution 
function % of the nucleon 
are also presented.

\end{abstract}

\maketitle

%%%%%%%%%%%%%%%%%%%%%%%%%%%%%%%%%%%%%%%%%%%%
%% MAINMATTER
%%%%%%%%%%%%%%%%%%%%%%%%%%%%%%%%%%%%%%%%%%%%

\section{Introduction}

Understanding the structure of the nucleon
from QCD has been one of the central issues 
in hadron physics.
In particular, the strangeness content of the nucleon 
attracts a great deal of interest lately.
It is also an ideal probe
for
the virtual sea quarks in the nucleon.
Extensive
experimental/theoretical studies indicate that
the strangeness content varies % significantly 
depending 
on the quantum number carried by the $s\bar{s}$ pair:
the % strangeness 
scalar density 
is about $0$--$20$\% 
of that of up, down quarks,
the quark spin is about $-10$ to $0$\%
of the nucleon,
and 
the momentum fraction is only 
a few percent of the nucleon.
In general, the uncertainties in the strangeness matrix elements
are quite large in both experiments and theories.
Under these circumstances,
it is desirable 
to provide the definitive quantitative results
using lattice QCD. %method.

The challenge in the
lattice QCD calculation of strangeness 
matrix elements 
resides in 
the evaluation of the so-called 
disconnected insertion (DI).
In fact, 
it requires the
calculation of 
all-to-all propagators,
which is prohibitively expensive
compared to the 
%calculation of the 
connected insertion (CI).
Consequently, there are only a few DI 
calculations~\cite{smm:ky_quenched,smm:ky_quenched2,smm:randy}, 
where the all-to-all propagators are stochastically estimated~\cite{DI:noise}.
In this proceeding,
we report the improvement of the calculation of 
all-to-all propagators 
using the stochastic method along with
unbiased subtractions from the
hopping parameter expansion~\cite{DI:hpe},
and the increment of the number of nucleon sources~\cite{x:deka,smm:doi}.
We present the results for the 
strangeness contribution to the 
electromagnetic form factors~\cite{smm:doi}
and the second moment of the nucleon.
The preliminary result for the
first moment of the nucleon is presented in
Ref.~\cite{x:doi}.

\section{Formalism and simulation parameters}

We employ %the 
$N_f=2+1$ dynamical 
configurations
with nonperturbatively ${\cal O}(a)$ improved clover fermion 
and 
%renormalization group 
RG-improved gauge action
generated by CP-PACS/JLQCD Collaborations~\cite{conf:tsukuba2+1}.
We use %$(\beta,c_{sw})=(1.83,1.761)$
$\beta=1.83$ and
$c_{sw}=1.7610$ 
configurations 
with the lattice size of $L^3 \times T = 16^3\times 32$,
which corresponds to
$(2{\rm fm})^3$ box in physical spacial size
with 
the lattice spacing of 
$a^{-1} = 1.625 {\rm GeV}$~\cite{conf:tsukuba2+1}.
For the hopping parameters of $u$,$d$ quarks ($\kappa_{ud}$) and
$s$ quark ($\kappa_s$), we use
$\kappa_{ud} =$ $0.13825$, $0.13800$, and $0.13760$,
which correspond to $m_\pi =$ $0.60$, $0.70$, and $0.84$ ${\rm GeV}$,
respectively, and $\kappa_s = 0.13760$ is fixed.
We perform the calculation only at the dynamical quark mass points,
where
800 configurations are used for $\kappa_{ud} =$ $0.13760$,
and 810 configurations % are used 
for
$\kappa_{ud} =$ $0.13800$, $0.13825$. % calculations.

The nucleon matrix elements can be obtained 
through the calculation of 3pt function $\Pi^{\rm 3pt}_{J}$
(as well as 2pt function $\Pi^{\rm 2pt}$),
defined by
\begin{eqnarray}
\Pi^{\rm 3pt}_{J}(\vec{p},t_2;\ \vec{q},t_1;\ \vec{p'}=\vec{p}-\vec{q},t_0)
&=&
\sum_{\vec{x_2},\vec{x_1}}
e^{-i\vec{p}\cdot(\vec{x}_2-\vec{x}_0)}
\cdot
e^{+i\vec{q}\cdot(\vec{x}_1-\vec{x}_0)}
\braket{0|{\rm T}\left[
\chi_N(\vec{x}_2,t_2) {J}(\vec{x}_1,t_1) \bar{\chi}_N(\vec{x}_0,t_0)
\right] |0}  ,
\label{eq:3pt}
\end{eqnarray}
where $\chi_N$ is the nucleon interpolating field
and $J$ is the insertion operator.
Since there is no strange quark as a valence quark in the nucleon,
%the 3pt amounts to the multiplication of DI and 2pt.
the 3pt is a DI which entails a multiplication of the nucleon 2pt correlator
with the current quark loop.
For the evaluation of the quark loop, 
we use the stochastic method~\cite{DI:noise},
with Z(4) noises in color, spin and space-time indices.
We generate independent noises for different configurations,
in order to avoid possible auto-correlation.
We use $N_{noise} = 600$ noises for $\kappa_{ud} = 0.13760, 0.13800$
and $N_{noise}=800$ for $\kappa_{ud} = 0.13825$. % calculations.
To reduce fluctuations, 
the charge conjugation and $\gamma_5$-hermiticity (CH), 
and parity symmetry are used~\cite{x:deka,smm:doi}.
%For instance, we find that the information for the $G_M^s$ is coded
%in the product of ${\rm Re}(\Pi^{\rm 2pt}) \times {\rm Re}({\rm loop})$,
%and filtering out the imaginary parts reduces the noises.
%
%
We also perform unbiased subtractions~\cite{DI:hpe} to 
reduce
the off-diagonal contaminations to the variance.
For subtraction operators,
we employ those obtained through
hopping parameter expansion (HPE) for the propagator $M^{-1}$,
%
%\begin{eqnarray}
$
\frac{1}{2\kappa} M^{-1} = \frac{1}{1+C} + \frac{1}{1+C} (\kappa D) \frac{1}{1+C} 
+ \cdots
%+  \frac{1}{1+C} (\kappa D) \frac{1}{1+C} (\kappa D) \frac{1}{1+C} + \cdots
%
$
%\end{eqnarray}
%
where $D$ denotes the Wilson-Dirac operator and $C$ the clover term.
We subtract up to order $(\kappa D)^4$ ($(\kappa D)^3$) term
for the form factor (second moment) calculation,
and observe that 
the statistical errors become
about 50 (70) \%,
compare to the results without subtraction.

In the stochastic method, 
it is quite expensive
to achieve a good signal to noise ratio (S/N) just by increasing $N_{noise}$
because S/N improves with $\sqrt{N_{noise}}$.
In view of this, we use many nucleon point sources $N_{src}$ 
in the evaluation of the 2pt part for each configuration.
Since the calculations of the loop part and 2pt part
are independent of each other, this is expected to be an efficient way.
%In particular, %for the $N_{noise} \gg N_{src}$ case,
%we observe that S/N improves almost ideally, by a factor of $\sqrt{N_{src}}$.
%
%
We take $N_{src}=64$ for $\kappa_{ud} = 0.13760$
and $N_{src}=82$ for $\kappa_{ud} = 0.13800, 0.13825$, % calculations,
where locations of sources are taken 
so that they are 
% dispersed 
separated
in 4D-volume as much as possible.
Details of the simulation setup are given in Ref.~\cite{smm:doi}.

%%%%%%%%%%%%%%%%%%%%%%%%%%%%%%%%%%%%%%%%%%%%%%%%%%%%%
%\newpage
\section{Strangeness electromagnetic form factors}

\begin{figure}[b]
\begin{minipage}{0.47\textwidth}
\begin{center}
\vspace*{-8mm}
\includegraphics[width=0.7\textwidth,angle=270]{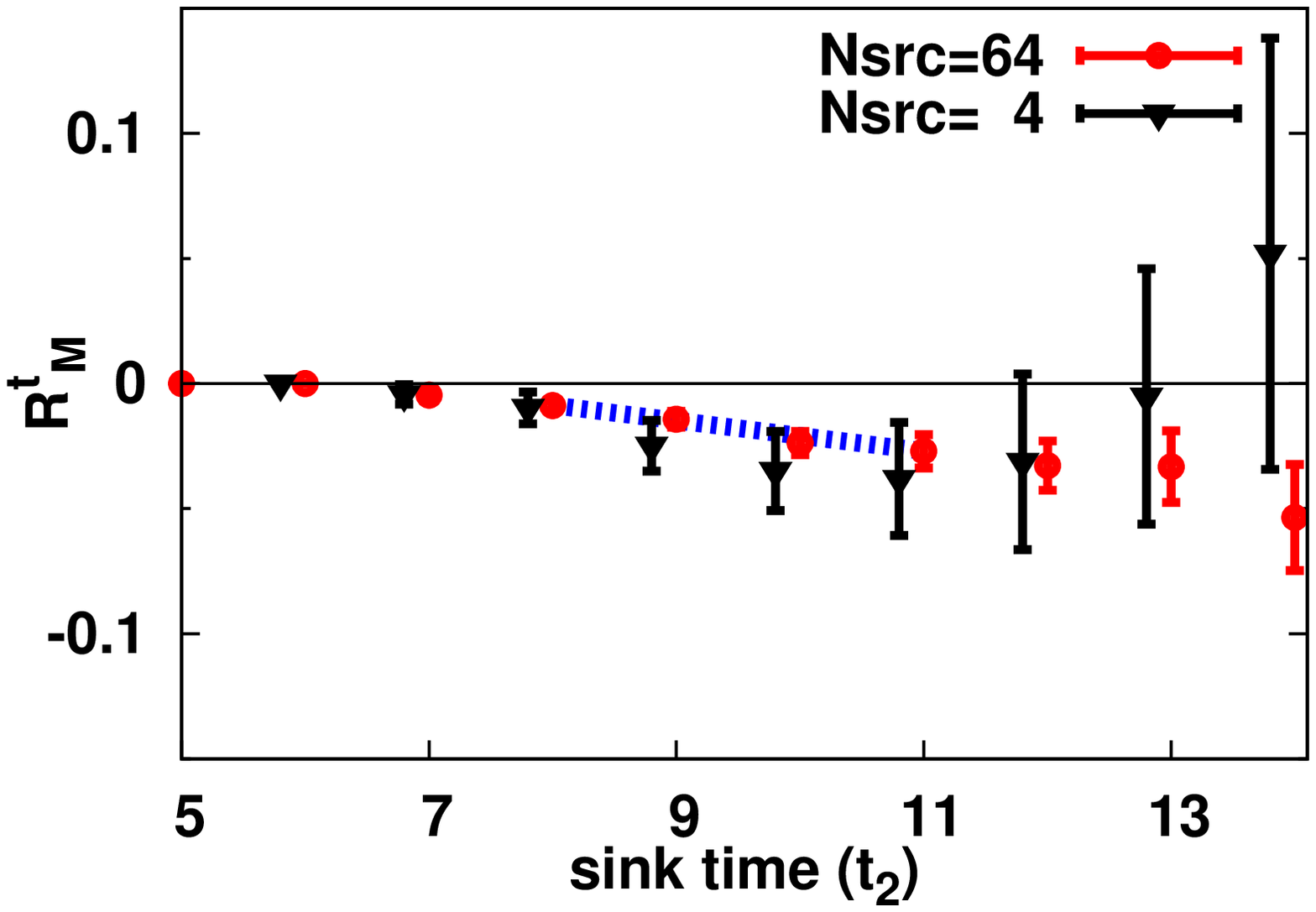}
\end{center}
\end{minipage}
\hfill
\begin{minipage}{0.47\textwidth}
\begin{center}
\vspace*{-8mm}
\includegraphics[width=0.7\textwidth,angle=270]{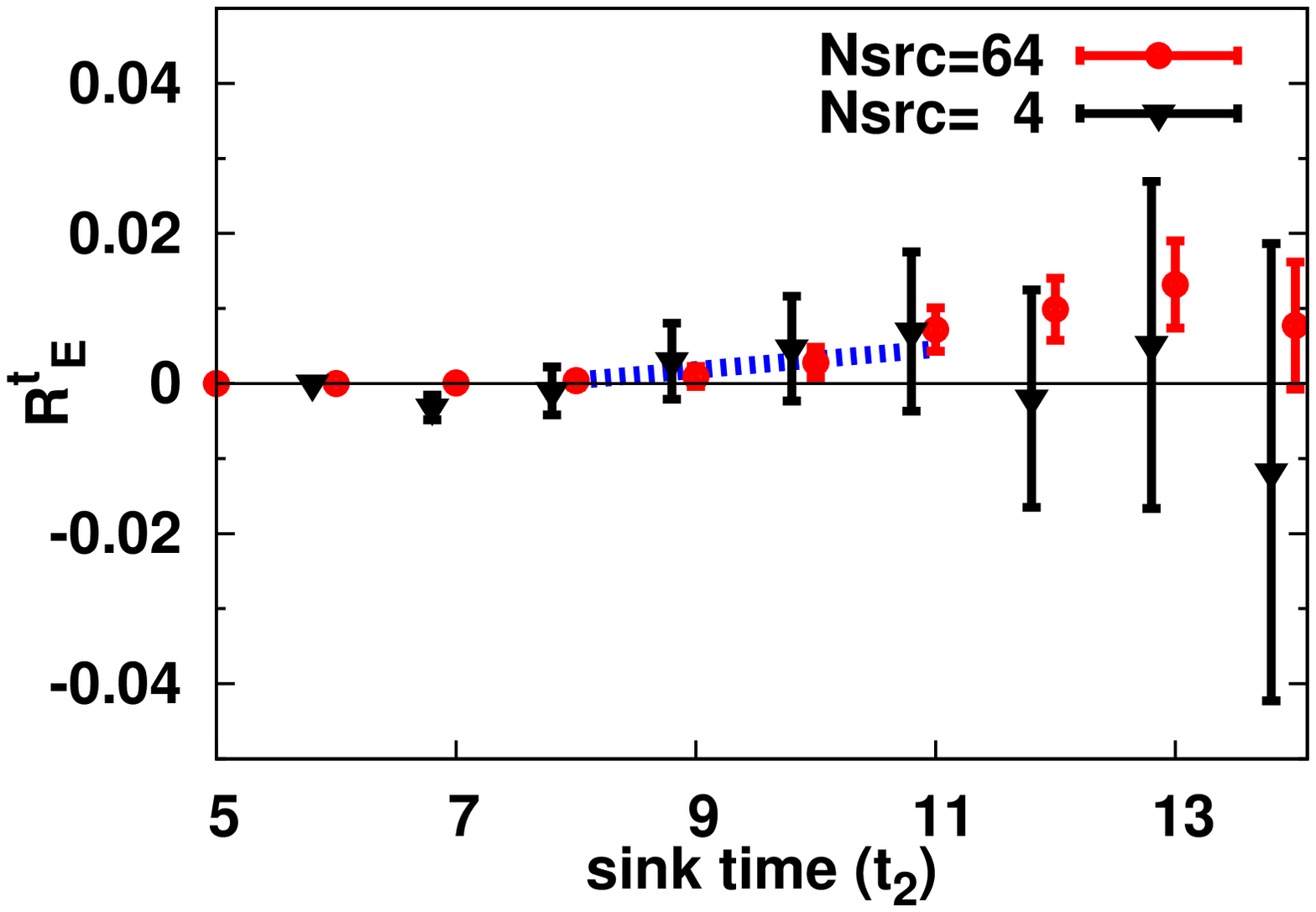}
%\hspace*{15mm}
\caption{
 \label{fig:tsum}
$R_M^t$ (left) and $R_E^t$ (right) with $\kappa_{ud} = 0.13760$,
%$\vec{q}^{\,2}= (2\pi/La)^2$,
$N_{src}=64$ (circles) and $N_{src}=4$ (triangles),
plotted against the nucleon sink time $t_2$.
The dashed line is the linear fit
where the slope corresponds to the form factor.
}
\end{center}
\end{minipage}
\end{figure}

The formulas for Sachs electric (magnetic) form factors
$G_E^s$ ($G_M^s$) 
are given by 
\begin{eqnarray}
%
%
%
%\lefteqn{
%
R_\mu^{\pm} (\Gamma_{\rm pol}^\pm) \equiv
\frac{
{\rm Tr}\left[
\Gamma_{\rm pol}^\pm\cdot
\Pi^{\rm 3pt}_{J_{\mu}}(\vec{0},{t}_2;\ \pm\vec{q},{t}_1;\ -\vec{q},t_0)
\right]
}
{
{\rm Tr} \left[
\Gamma_e^\pm\cdot \Pi^{\rm 2pt}_{}(\pm\vec{q},t_1;\ t_0)
\right]
}
%
%} 
%\nn
%
%&&
%
\cdot
%\qquad
%\times
\frac{
{\rm Tr} \left[
\Gamma_e^\pm\cdot \Pi^{\rm 2pt}_{}(\vec{0},t_1;\ t_0)
\right]
}
{
{\rm Tr} \left[
\Gamma_e^\pm\cdot \Pi^{\rm 2pt}_{}(\vec{0},t_2;\ t_0)
\right]
} , \\
\label{eq:ratio,kin1,fb}
%
%
%
%\nn[-8mm]
%\end{eqnarray}
%
%\vspace*{-5mm}
%
%\begin{eqnarray}
%
%
G_E^s(Q^2) = \pm R_{\mu=4}^{\pm}(\Gamma_{\rm pol}^\pm=\Gamma_e^\pm) ,
\label{eq:ele1,kin1,fb} 
\qquad
G_M^s(Q^2) = 
\mp
\frac{E^q_N+m_N}{\epsilon_{ijk} q_j}
R_{\mu=i}^{\pm}(\Gamma_{\rm pol}^\pm=\Gamma_k^\pm) , 
\label{eq:mag,kin1,fb}
%
%
%\nn[-8mm]
\end{eqnarray}
where 
$
J_\mu (x+\mu/2)
=
\frac{1}{2}
\left[
\bar{q}(x) (1-\gamma_\mu) U_\mu(x) q(x+\mu)
- \bar{q}(x+\mu) (1+\gamma_\mu) U_\mu^\dag(x) q(x)
\right]
$ 
is the point-split conserved vector operator,
$\{i,j,k\} \neq 4$,
$
\Gamma_e^\pm \equiv (1\pm\gamma_4)/2
$ ,
$
\Gamma_k^\pm \equiv (\pm i)/2 \times (1\pm\gamma_4) \gf \gamma_k
$ 
and
%
%$m_N$ denotes the nucleon mass, 
$E^q_N \equiv \sqrt{m_N^2 + \vec{q}^{\,2}}$.
%
%
%
%For the choice of double sign,
%The forward propagation corresponds to the upper sign
%and the backward % propagation corresponds 
%to the lower sign.
The upper sign corresponds to 
the forward propagation ($t_2 \gg t_1 \gg t_0$),
and the lower sign corresponds to 
the backward propagation ($t_2 \ll t_1 \ll t_0$).

In Fig.~\ref{fig:tsum},
we plot
typical figures
for $R^{t}_{M,E}$, where
$R^{t}_{M,E} \equiv 
\frac{1}{K_{M,E}^\pm}
\sum_{t_1=t_0+t_{s}}^{t_2-t_{s}}
R^{\pm}_\mu
$
with $K_{M,E}^\pm$ being trivial kinematic factors
%appearing 
in Eq.~(\ref{eq:ele1,kin1,fb}). % and (\ref{eq:mag,kin1,fb}),
Since 
$R^{t}_{M,E}
=
{\rm const.} + t_2 \times G_{M,E}^s ,
$
the linear slope corresponds to the signal of the form factor.
One can 
observe 
the significant S/N improvement by increasing $N_{src}$.
In fact, the improvement is found to be nearly
a factor of $\sqrt{N_{src}}$ (ideal improvement).

We then study the 
$Q^2$ dependence
of the form factors.
For the magnetic form factor,
we employ the dipole form,
$G_M^s(Q^2) = G_M^s(0) / (1+Q^2/\Lambda^2)^2$,
where reasonable agreement with lattice data is observed.
%where correlations among 
%different $Q^2$ are taken into account.
%
%
%
For the electric form factor,
we employ 
$G_E^s(Q^2) = g_E^s \cdot Q^2 / (1+Q^2/\Lambda^2)^2$,
considering that $G_E^s(0)=0$ from the vector current 
conservation, and the pole mass $\Lambda$ is taken from
the fit of magnetic form factor.
%We also test the simultaneous fit of 
%$G_M^s(Q^2)$ and $G_E^s(Q^2)$ with 
%three parameters of $G_M^s(0), \Lambda, g_E^s$, %in total,
%and confirm that the results are consistent.

Finally, we perform the chiral extrapolation for the fitted parameters.
Since our quark masses are relatively heavy, 
we consider only the leading dependence on $m_K$,
which is obtained by heavy baryon chiral perturbation theory (HB$\chi$PT)%
~\cite{chPT:hemmert}.
%For the magnetic moment $G_M^s(0)$, we fit linearly in terms of $m_K$.
%For the pole mass $\Lambda$, we take that the magnetic mean-square radius
%$\braket{r^2_s}_M \equiv -6 \frac{d G_M^s}{d Q^2}|_{Q^2=0} = 12 G_M^s(0) / \Lambda^2$
%behaves as $1/m_K$.
%For $g_E^s$, we use the electric radius
%$\braket{r^2_s}_E \equiv -6 \frac{d G_E^s}{d Q^2}|_{Q^2=0} = -6 g_E^s$
%which has an $\ln (m_K/\mu)$ behavior, and we take the scale $\mu=1$ GeV.
%
%
%
The chiral extrapolated results are
$G_M^s(0) = -0.017(25)$,
$\Lambda a = 0.58(16)$,
$\braket{r_s^2}_M 
\equiv -6 \frac{d G_M^s}{d Q^2}|_{Q^2=0} 
= -7.4(71) \times 10^{-3} {\rm fm}^2$
and
$g_E^s = 0.027(16)$
(or $\braket{r_s^2}_E
\equiv -6 \frac{d G_E^s}{d Q^2}|_{Q^2=0} 
= -2.4(15) \times 10^{-3} {\rm fm}^2$).

We examine the systematic uncertainties in the result of form factors.
For the ambiguity of $Q^2$ dependence, % in form factors,
we reanalyze the data using the monopole form,
and obtain the results which are consistent with
those from the dipole form.
For the uncertainties in chiral extrapolation,
we test two alternative extrapolations~\cite{smm:doi},
and find that all results are consistent with each other.
For the contamination from excited states,
we employ the new projection operator~\cite{smm:doi}
which eliminates the 
%contamination from 
$S_{11}$ state,
and conclude that such contaminations are negligible.

\begin{figure}[tb]
%\begin{center}
%
%\hspace*{-10mm}
\includegraphics[width=0.35\textwidth,angle=270]{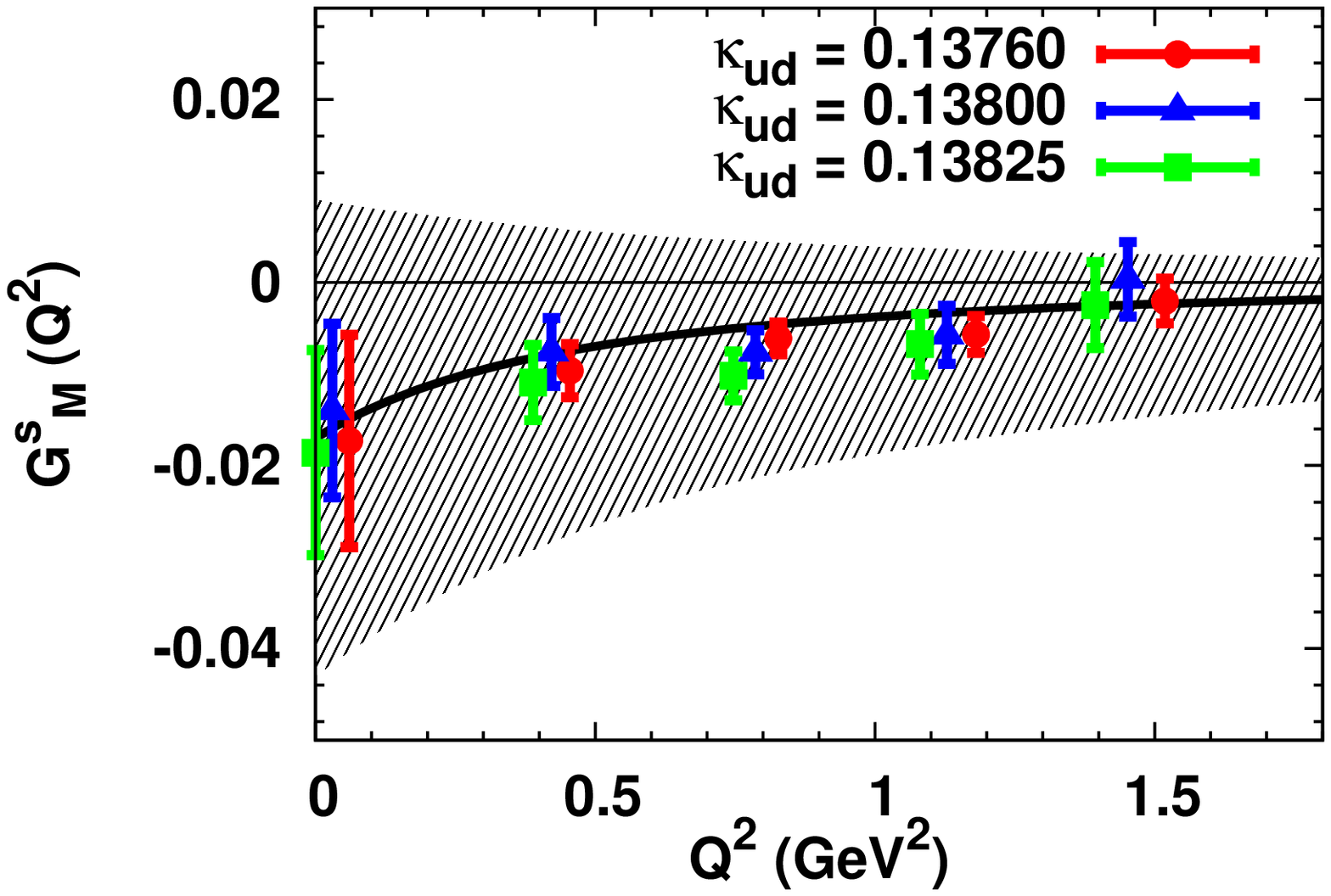} %\\[-3mm]
%\vspace*{-5mm}
%\hspace*{-10mm}
\includegraphics[width=0.35\textwidth,angle=270]{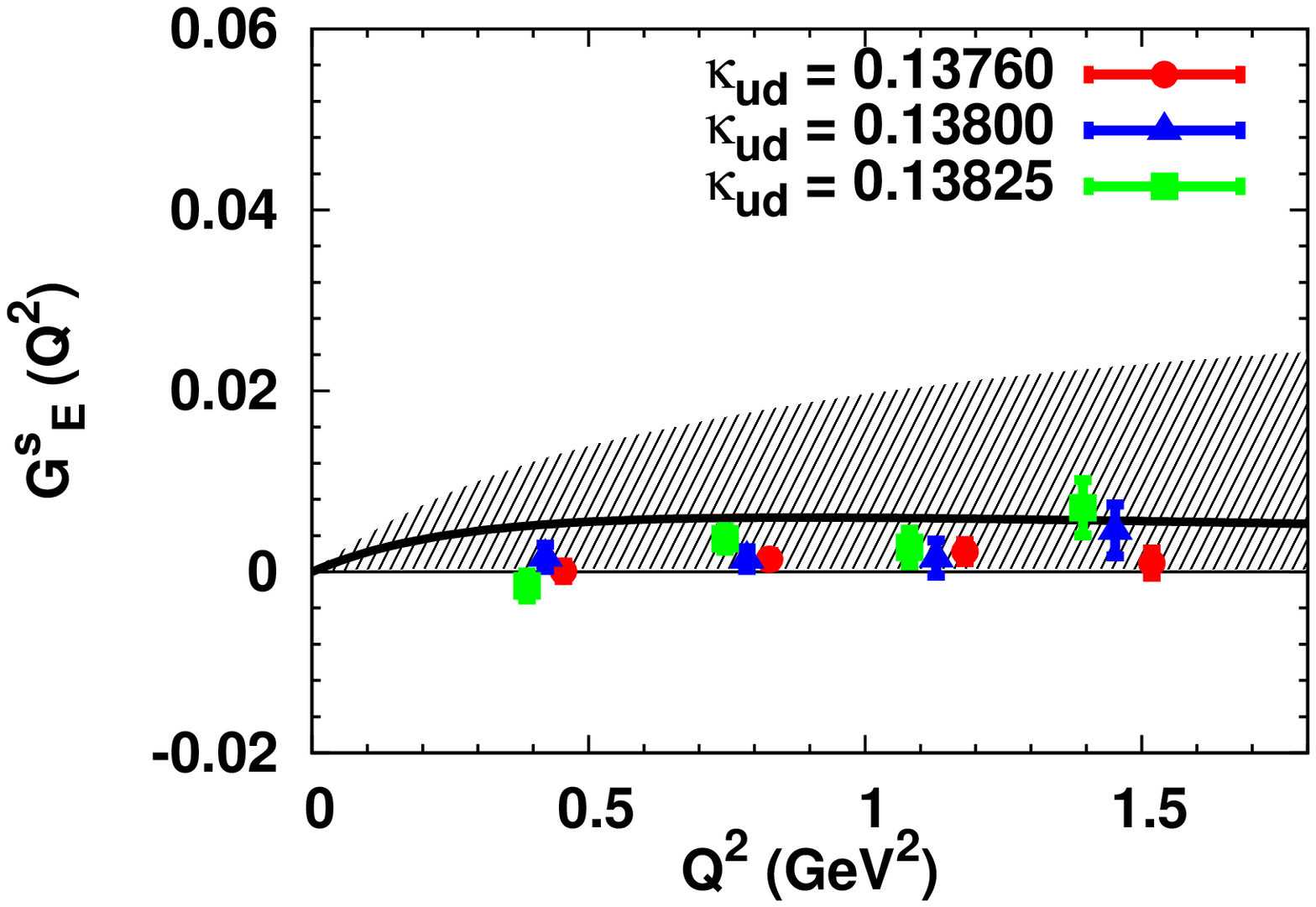}
 \label{fig:res}
%\end{center}
\caption{
The chiral extrapolated results for $G_M^s(Q^2)$ (left) 
and $G_E^s(Q^2)$ (right) plotted with solid lines.
Shaded regions 
represent
%the one-$\sigma$ uncertainty 
%the error-band 
%with 
the
statistical and systematic error
added in quadrature.
%Plotted 
Shown
together are the lattice data
for each $\kappa_{ud}$.
%(and $Q^2$-extrapolated $G_M^s(0)$)
%for
%$\kappa_{ud}=$ 
%0.13760 (circles),
%0.13800 (triangles),
%0.13825 (squares)
%with offset for visibility.
%For the $G_M^s(Q^2)$ figure, the $Q^2$ extrapolated 
%data for each $\kappa_{ud}=$ are also shown.
}
\end{figure}

Our final result for the magnetic moment is 
$G_M^s(0) = -0.017(25)(07)$, 
where the first error is statistical and
the second is systematic %error 
from uncertainties of the $Q^2$ extrapolation and chiral extrapolation.
We also obtain $\Lambda a = 0.58(16)(19)$ for dipole mass
or $\tilde{\Lambda} a = 0.34(17)(11)$ for monopole mass,
and $g_E^s = 0.027(16)(08)$.
These lead to 
%from which we obtain 
$G_M^s(Q^2) = -0.015(23)$,
$G_E^s(Q^2) =  0.0022(19)$ at $Q^2=0.1 {\rm GeV}^2$,
where error is obtained by quadrature from 
statistical and systematic errors.
In Fig.~\ref{fig:res}, 
we plot 
$G_M^s(Q^2)$, $G_E^s(Q^2)$,
where the shaded regions 
correspond to the square-summed error.
%The lattice data for each $\kappa_{ud}$
%are also plotted.
%
%
%
Compared to the global analysis
of the experimental data,
e.g., 
$G_M^s(Q^2) = 0.29(21)$ and
$G_E^s(Q^2) = -0.008(16)$ 
at $Q^2 = 0.1 {\rm GeV}^2$~\cite{pate08},
our results are consistent with them,
with an order of magnitude smaller error~\cite{smm:doi}.

%%%%%%%%%%%%%%%%%%%%%%%%%%%%%%%%%%%%%%%%%%%%%%%%%%%%%
%\newpage
\section{Second moment of the nucleon}

\begin{figure}[t]
\begin{minipage}{0.47\textwidth}
 \begin{center}
   \includegraphics*[width=0.7 \textwidth,angle=270]
   {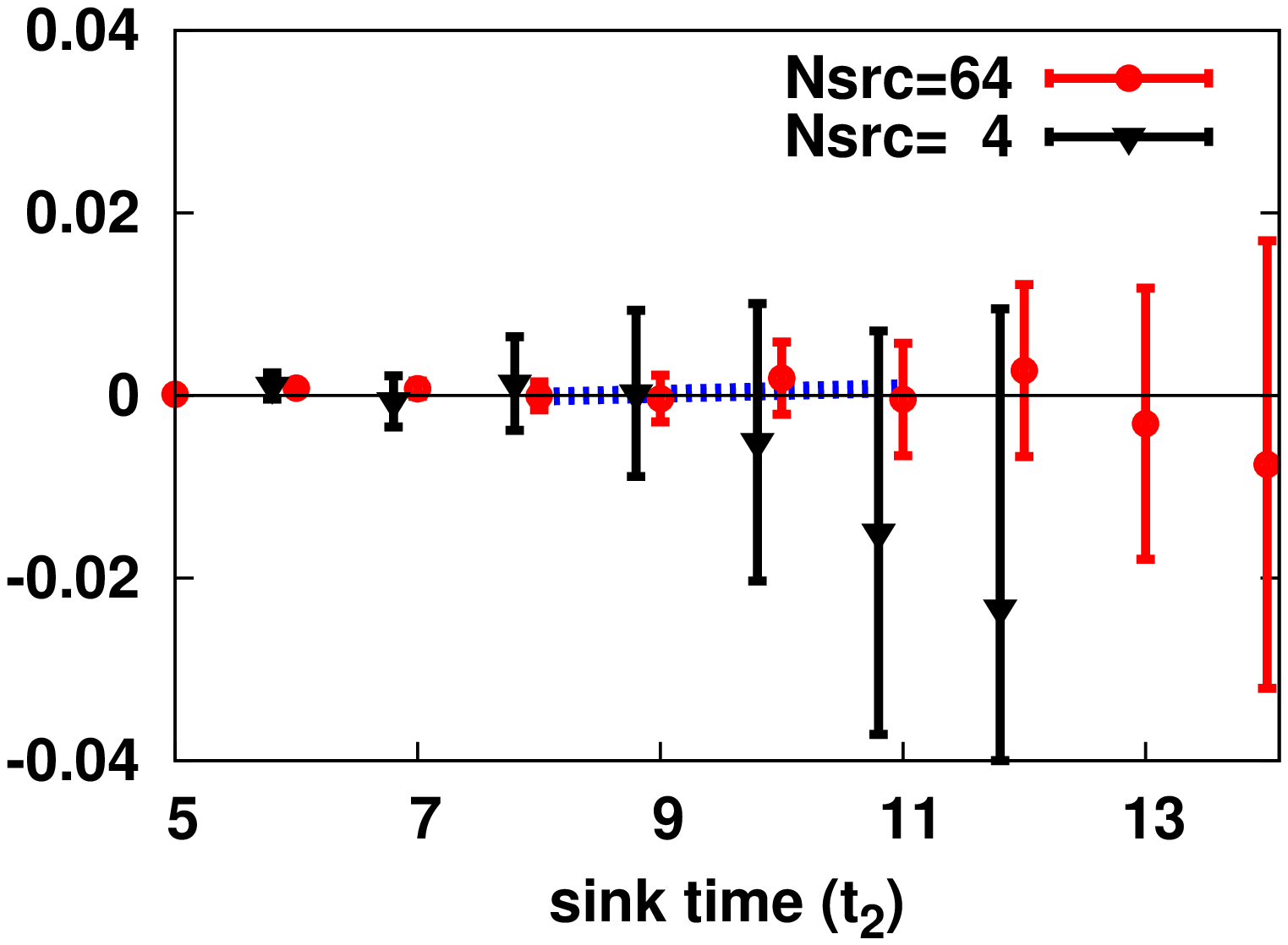}
 \caption{}
 \end{center}
\end{minipage}
\hfill
\begin{minipage}{0.47\textwidth}
 \begin{center}
   \includegraphics*[width=0.7 \textwidth,angle=270]
   {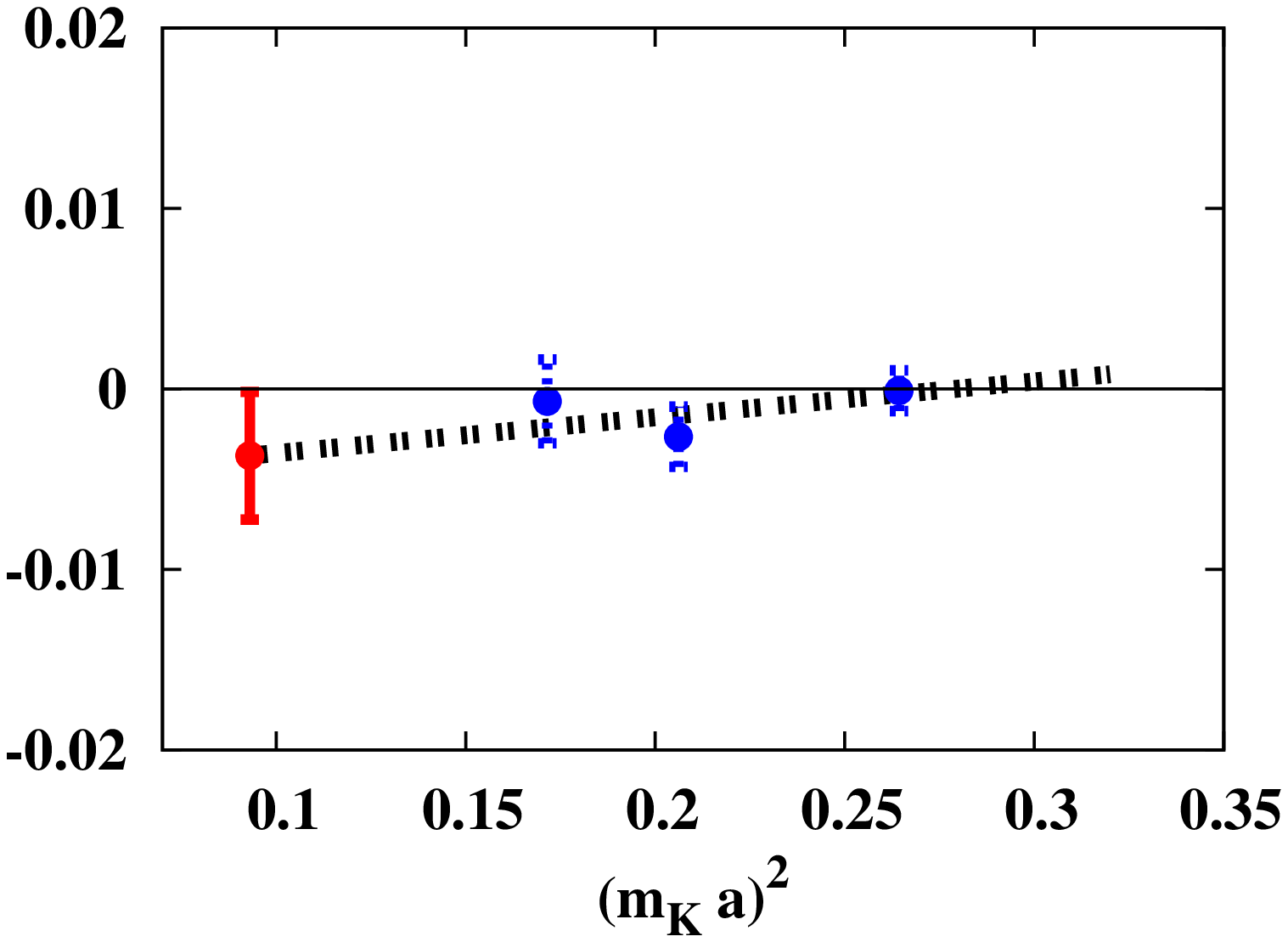}
   \caption{}
 \end{center}
\end{minipage}
 \caption{
 \label{fig:xx_tsum}
LEFT:
The ratio of 3pt to 2pt with $\kappa_{ud} = 0.13760$,
%$\vec{q}^{\,2}=2\cdot (2\pi/La)^2$,
$N_{src}=64$ (circles) and $N_{src}=4$ (triangles),
plotted against the nucleon sink time $t_2$.
The dashed line is the linear fit
where the slope corresponds to the 
second moment.
RIGHT:
The lattice bare results for the second moment
at each valence quark mass $\kappa_{ud}$ for the nucleon,
plotted against $(m_K a)^2$.
The dashed line corresponds to the linear chiral extrapolation,
and the red point is the chiral extrapolated result.
}
\end{figure}

The (asymmetry of) strangeness second moment of the nucleon 
$\langle x^2 \rangle_{s-\bar{s}} = \int_0^1 dx x^2 ( s(x) - \bar{s}(x) )$
can be obtained by

\begin{eqnarray}
\frac{
{\rm Tr}\left[
\Gamma^\pm_e \cdot
\Pi^{\rm 3pt}_{T_{4ii}}(\pm\vec{p},t_2;\ \vec{0},t_1;\ \pm\vec{p},t_0)
\right]
}
{
{\rm Tr} \left[
\Gamma^\pm_e \cdot \Pi^{\rm 2pt}(\pm\vec{p},t_2;\ t_0)
\right]
}
=
\pm p_i^2 \cdot \xx_{s-\bar{s}}, 
\end{eqnarray}
with 
the three-index operator defined as
\begin{eqnarray}
T_{4ii}
\equiv
-\frac{1}{3}
\left[ 
  \bar{q} \gamma_4 \overleftrightarrow{D}_i \overleftrightarrow{D}_i q
+ \bar{q} \gamma_i \overleftrightarrow{D}_4 \overleftrightarrow{D}_i q
+ \bar{q} \gamma_i \overleftrightarrow{D}_i \overleftrightarrow{D}_4 q
\right] ,
\end{eqnarray}
where $i\neq 4$,
and the upper (lower) sign corresponds to the forward (backward)
propagation as before.

In Fig. \ref{fig:xx_tsum} (left), we plot the ratio of 3pt to 2pt 
for $\xx_{s-\bar{s}}$ 
in terms of $t_2$ for $\kappa_{ud} = 0.13760$, $\vec{p}^{\,2}=(2\pi/La)^2$,
where the summation of operator insertion time $t_1$ is taken
as was done for the form factor analysis.
Note that the linear slope corresponds to the signal for $\xx_{s-\bar{s}}$.
%Blue points denote the result for $N_{src}=4$ and red points
%denote $N_{src}=64$.
One can clearly see that increasing $N_{src}$ reduces the error bar
significantly again (about a factor of $\sqrt{N_{src}}$, i.e., almost ideally).
In Fig. \ref{fig:xx_tsum} (right), 
we plot the bare value of the $\xx_{s-\bar{s}}$
in terms of $(m_K a)^2$,
and perform the chiral extrapolation.
We find that 
the result at each $\kappa_{ud}$ 
and the chiral extrapolated result
are basically consistent with zero 
within the error-bar.
For the final quantitative result, it is necessary to take
the renormalization factor into account.
Systematic uncertainties have to be examined as well.
The study along this line is in progress.

%\section{Summary}

%In summary, 
%we have studied the strangeness electromagnetic form factors
%of the nucleon from the $N_f=2+1$ clover fermion lattice QCD calculation.
%It has been found that calculating many nucleon sources is essential to achieve 
%a good S/N in the evaluation of DI.
%We have obtained the form factors 
%which are consistent with experimental values,
%and have an order of magnitude smaller error.
%A preliminary result for the strangeness second moment, $\xx_{s-\bar{s}}$
%has been reported.

%%%%%%%%%%%%%%%%%%%%%%%%%%%%%%%%%%%%%%%%%%%%%%%%
%% BACKMATTER
%%%%%%%%%%%%%%%%%%%%%%%%%%%%%%%%%%%%%%%%%%%%%%%%

\begin{theacknowledgments}

We thank the CP-PACS/JLQCD Collaborations
for their configurations.
This work was supported  in part by 
U.S. DOE grant DE-FG05-84ER40154.
TD is supported in part by
Grant-in-Aid for JSPS Fellows 21$\cdot$5985.
Research of NM is supported by Ramanujan Fellowship.
The calculation was performed 
at Jefferson Lab, Fermilab
and 
the University of Kentucky,
partly using the Chroma Library~\cite{chroma}.

\end{theacknowledgments}

%%%%%%%%%%%%%%%%%%%%%%%%%%%%%%%%%%%%%%%%%%%%%%%%
%% The bibliography can be prepared using the BibTeX program or
%% manually.
%%
%% The code below assumes that BibTeX is used.  If the bibliography is
%% produced without BibTeX comment out the following lines and see the
%% aipguide.pdf for further information.
%%
%% For your convenience a manually coded example is appended
%% after the \end{document}
%%%%%%%%%%%%%%%%%%%%%%%%%%%%%%%%%%%%%%%%%%%%%%%%

%%%%%%%%%%%%%%%%%%%%%%%%%%%%%%%%%%%%%%%%%%%%%%%%
%% You may have to change the BibTeX style below, depending on your
%% setup or preferences.
%%
%%
%% For The AIP proceedings layouts use either
%%%%%%%%%%%%%%%%%%%%%%%%%%%%%%%%%%%%%%%%%%%%

\bibliographystyle{aipproc}   % if natbib is available
%\bibliographystyle{aipprocl} % if natbib is missing

%%%%%%%%%%%%%%%%%%%%%%%%%%%%%%%%%%%%%%%%%%%
%% You probably want to use your own bibtex database here
%%%%%%%%%%%%%%%%%%%%%%%%%%%%%%%%%%%%%%%%%%%
%\bibliography{sample}

%%%%%%%%%%%%%%%%%%%%%%%%%%%%%%%%%%%%%%%%%%%
%% Just a reminder that you may have to run bibtex
%% All of it up to \end{document} can be removed
%% if you don't like the warning.
%%%%%%%%%%%%%%%%%%%%%%%%%%%%%%%%%%%%%%%%%%%
%\IfFileExists{\jobname.bbl}{}
% {\typeout{}
%  \typeout{******************************************}
%  \typeout{** Please run "bibtex \jobname" to optain}
%  \typeout{** the bibliography and then re-run LaTeX}
%  \typeout{** twice to fix the references!}
%  \typeout{******************************************}
%  \typeout{}
% }

%\end{document}

%%%%%%%%%%%%%%%%%%%%%%%%%%%%%%%%%%%%%%%%%%%
%% The following lines show an example how to produce a bibliography
%% without the help of the BibTeX program. This could be used instead
%% of the above.
%%%%%%%%%%%%%%%%%%%%%%%%%%%%%%%%%%%%%%%%%%%

\end{document}